\documentclass[aps,prl,onecolumn,superscriptaddress,groupedaddress]{revtex4}
\usepackage{subfig}
\usepackage{graphicx}  
\usepackage{dcolumn}   
\usepackage{bm}        
\usepackage{amssymb}   
\usepackage{slashed}
\usepackage{graphicx}				
\usepackage{amsmath}
\usepackage{mathtools}
\usepackage{tikz,pgf}
\usepackage{comment}
\usepackage{asymptote}
\usetikzlibrary{arrows,backgrounds}
\usetikzlibrary{fit,scopes,calc,matrix,positioning,decorations.pathmorphing}
\usepackage[all]{xy}
\usepackage{yfonts}

\newcommand{\Bracket}[1]{\ensuremath{\left\langle#1\right\rangle}}

\begin{document}
\title{Structured Nonlinear Cascades Bridging Macroscopic Fluid Scales and Molecular Vibrations}
\author{Andrei T. Patrascu}
\address{FAST Foundation, Destin FL, 32541, USA\\
email: andrei.patrascu.11@alumni.ucl.ac.uk}
\begin{abstract}
We propose and theoretically analyze a novel approach to selectively excite molecular vibrational modes through structured fluid dynamics guided by generalized symmetry-based transformations of the Navier–Stokes equations. By encoding specific molecular resonance information into structured macroscopic fluid perturbations and using iterative nonlinear cascades, we demonstrate numerically that energy can coherently transfer from macroscopic scales down to molecular vibrational frequencies. This structured cascade, described by a generalized Gelfand transform and associated nonlinear structure constants, ensures resonance conditions at molecular scales, significantly delaying thermalization and enabling precise quantum state manipulation in fluids. Numerical simulations explicitly targeting the asymmetric vibrational mode of $CO_{2}$ validate this methodology, highlighting its potential applications in controlled molecular excitation and coherent fluid-based quantum manipulation.
\end{abstract}
\maketitle
\section{Introduction}
Precise excitation and control of molecular quantum states through macroscopic fluid dynamics [1], [2], [3] remains a formidable challenge due to the vast disparity in frequency scales between macroscopic fluid motions and molecular vibrations. Traditional fluid turbulence cascades energy chaotically down to microscopic scales, rapidly dissipating structured information and limiting coherent molecular state manipulation [4], [5], [6]. However, the controlled and coherent excitation of specific molecular vibrational modes would unlock significant potential in chemical reaction control, quantum technology, and energy harvesting applications.

In this Letter, we introduce a general theoretical framework based on generalized Gelfand transformations [7] of the Navier–Stokes equations [8], explicitly incorporating molecular symmetry into structured fluid modes. This structured approach utilizes nonlinear interactions and iterative cascades across multiple scales to progressively guide energy toward precise molecular resonances. Numerical simulations demonstrate the feasibility of selectively exciting the asymmetric stretch mode of the $CO_{2}$ molecule, illustrating the effectiveness of symmetry-based fluid perturbations. Our approach reveals that structured symmetry-preserving cascades can efficiently bridge macroscopic-to-molecular scales, overcoming conventional turbulence limitations and enabling a new class of coherent fluid-based quantum manipulations.

\section{Theoretical Framework}

The fundamental idea behind our structured fluid cascade approach is the generalisation of the Navier–Stokes equations using the Gelfand transform.
The central idea is to start with an initial fluid perturbation structured by symmetry to resonate with a targeted molecular mode. Then we solve the Navier Stokes equations numerically, in a symmetry adapted form (via Gelfand transforms) and use the computed solution as feedback to adjust the initial structured perturbation iteratively, achieving increasingly accurate resonant conditions at the molecular scale. 
 This method begins by expressing the classical Navier–Stokes equations in terms of fluid velocity fields:
\begin{equation}
\rho\frac{\partial u}{\partial t}+\rho(u\cdot\nabla)u=-\nabla P+\mu\nabla^{2}u +\rho f
\end{equation}

where $\rho$ is the fluid density, $u$ is the velocity field, $P$ is the pressure, $\mu$  is viscosity, and $f$ is the external forcing.

To systematically encode molecular symmetry and resonance information into fluid dynamics, we apply the Gelfand transform, a group-theoretic generalization of the Fourier transform.
For abelian groups, the Gelfand transform reduces to the standard Fourier transform, decomposing functions into exponential modes indexed by characters of the group. In the non-abelian case, however, irreducible unitary representations are generally multidimensional, resulting in matrix-valued modes that encode richer symmetry structures.
Specifically, the fluid velocity field is decomposed into structured symmetry modes labeled by irreducible representations $\pi$ of the molecular symmetry group $G$ :
\begin{equation}
\hat{u}(\pi,t)=\int_{G} u(g,t)\pi(g^{-1})dg
\end{equation}
where $g\in G$, and the integration is over the molecular symmetry group.
The inverse Gelfand transform reconstructs the velocity field as 
\begin{equation}
u(g,t)=\sum_{\pi}dim(\pi)Tr[\hat{u}(\pi,t)\pi(g)]
\end{equation}

This transform maps the Navier–Stokes equations into structured symmetry space, resulting in explicitly symmetry-adapted structured fluid modes. The nonlinear interaction term transforms into:
\begin{equation}
(u\cdot \nabla)u\rightarrow \sum_{\pi_{1},\pi_{2}}\Gamma(\pi,\pi_{1},\pi_{2})\hat{u}(\pi_{1},t)(\hat{\pi}_{2},t)
\end{equation}

where the structure constants $\Gamma(\pi, \pi_{1},\pi_{2})$ are obtained via symmetry-based integrals:
\begin{equation}
\Gamma(\pi, \pi_{1}, \pi_{2})=\int_{G}\pi(g^{-1})[(\pi_{1}(g)\cdot \nabla)\pi_{2}(g)]dg
\end{equation}

To describe multiple cascades explicitly, we employ an iterative numerical method. Initially, structured modes corresponding to lower frequency scales are excited by external perturbations. After numerically resolving the generalised Gelfand–Navier–Stokes equations at each intermediate scale, the resulting structured fields guide adjustments in experimental parameters such as geometry, forcing frequencies, and boundary conditions are implemented. These iterative adjustments are determined by analysing how nonlinear interactions shift energy distributions towards higher-frequency molecular resonances. 
The transition from a macroscopic fluid perturbation with a relatively low frequency, down to a molecular vibrational frequency would involve multiple intermediate cascades and non-linear frequency shifts across several scales. If we consider the macroscopic fluid perturbations typically in the range of $Hz$ to $kHz$ or $MHz$ frequencies (acoustic waves to ultrasound) and the molecular vibrational modes typically in the $THz$ range $(\sim 10^{12} Hz)$, where for example in the $CO_{2}$ case the asymmetric stretch mode is at 2349 $cm^{-1}$ which would be $\sim 70 THz$ the frequency gap would be $10^6-10^{12}$ $Hz$. Each nonlinear fluid cascade typically allows frequency multiplication on the order of a factor of approximately 2 to 10. Let us consider an average frequency multiplication factor of $f_{cascade}\sim 5$. The total number of cascades $N$ required would be $f_{cascade}^{N}\sim 10^{6}$ and hence
\begin{equation}
N \log_{10}(f_{cascade})\sim log_{10}(10^{6})
\end{equation}
Using the chosen cascade factor $(f_{cascade}=5)$ we would have 
\begin{equation}
N\log_{10}(5)\sim 6 \Rightarrow N\sim\frac{6}{\log_{10}(5)}\sim \frac{6}{0.699}\sim 8.6
\end{equation}
Thus, around $8$ to $9$ cascade steps are realistically required to bridge from $MHz$ scale to $THz$ scale. 
Realistic estimations indicate that bridging typical macroscopic frequencies ($\sim$MHz) to molecular vibrational frequencies ($\sim$THz) typically requires around 8-10 cascades, with each cascade step involving frequency multiplications by factors of about 5-10. This method ensures precise matching of symmetry and frequency conditions at each cascade step, progressively steering energy transfer to targeted molecular vibrational modes.

Explicitly, we follow this algorithm starting with the generalised Navier-Stokes equation written in the symmetry adapted form using the Gelfand transformation
\begin{equation}
\rho\frac{\partial \hat{u}(\pi,t)}{\partial t}+\rho\sum_{\pi_{1},\pi_{2}}\Gamma(\pi,\pi_{1},\pi_{2})\hat{u}(\pi_{1},t)\hat{u}(\pi_{2},t)=-\hat{P}(\pi,t)-\mu |\pi|^{2}\hat{u}(\pi,t)+\rho\hat{f}(\pi,t)
\end{equation}
where $\hat{u}(\pi,t)$ is the structured fluid velocity mode in the irreducible representation $\pi$, $\Gamma(\pi,\pi_{1},\pi_{2})$ are nonlinear coupling constants defined as symmetry based integrals in the way I showed previously, and $\hat{f}(\pi,t)$ is the structured forcing. To automate dynamical adjustments, we formulate the problem as an optimal control problem. We define the cost functional $J$ as
\begin{equation}
J[\hat{u}(\pi,t),\hat{f}(\pi,t)]=\frac{1}{2}\int_{0}^{T}\sum_{\pi}||\hat{u}(\pi,t)-\hat{u}^{target}(\pi)||^{2}dt+\frac{\lambda}{2}\int_{0}^{T}||\hat{f}(\pi,t)||^{2}dt
\end{equation}
The first term penalises deviations from the target structured mode $\hat{u}^{target}(\pi)$. The second term penalises the magnitude of adjustments (control effort) $\hat{f}(\pi,t)$ and $\lambda$ is a regularisation parameter controlling trade-off between accuracy and effort. To find the optimal structured adjustments $\hat{f}(\pi,t)$ we introduce a Lagrange multiplier (adjoint field) $\hat{v}(\pi,t)$ and define a Lagrangian functional 
\begin{equation}
\mathcal{L}=J[\hat{u},\hat{f}]+\int_{0}^{T}\sum_{\pi}\Bracket{\hat{v}(\pi,t),\rho\frac{\partial \hat{u}(\pi,t)}{\partial t}+\rho\sum_{\pi_{1},\pi_{2}}\Gamma(\pi,\pi_{1},\pi_{2})\hat{u}(\pi_{1},t)\hat{u}(\pi_{2},t)+\hat{P}(\pi,t)+\mu |\pi|^{2}\hat{u}(\pi,t)-\rho\hat{f}(\pi,t)}dt
\end{equation}
Taking the variations of $\mathcal{L}$ with respect to $\hat{u}(\pi,t)$, $\hat{f}(\pi,t)$, and $\hat{v}(\pi,t)$ gives 
the forward state equation, which is the original structured Navier-Stokes equation and the adjoint (backward) equation which comes from the variation with respect to $\hat{u}$
\begin{equation}
-\rho\frac{\partial\hat{v}(\pi,t)}{\partial t}+\rho\sum_{\pi_{1},\pi_{2}}\Gamma(\pi_{1},\pi,\pi_{2})[\hat{v}(\pi_{1},t)\hat{u}(\pi_{2},t)+\hat{u}(\pi_{1},t)\hat{v}(\pi_{2},t)]+\mu |pi|^{2}\hat{v}(\pi,t)=\hat{u}(\pi,t)-\hat{u}^{target}(\pi)
\end{equation}
The boundary condition at final time $T$ is $\hat{v}(\pi,T)=0$. The optimality condition (namely the variation with respect to $\hat{f}$ is 
\begin{equation}
\hat{f}(\pi,t)=\frac{1}{\lambda}\hat{v}(\pi,t)
\end{equation}
This leads to a numerical algorithm which has been implemented: the initialisation phase implies setting up an initial guess of $\hat{f}(\pi,t)$, identifying the target mode $\hat{u}^{target}(\pi)$, and setting up the initial condition $\hat{u}(\pi,0)$. The forward simulation involves solving the structured Navier Stokes forward equation to obtain $\hat{u}(\pi,t)$. Then we compute the cost $J[\hat{u},\hat{f}]$ and we perform a backward simulation solving the adjoint equation backward in time to obtain $\hat{v}(\pi,t)$. We update the control (make an adjustment) 
\begin{equation}
\hat{f}_{new}(\pi,t)=\frac{1}{\lambda}\hat{v}(\pi,t)
\end{equation}
and repeat the steps above until convergence. This results in an adjustment of the structured perturbations at each iteration to match the desired resonance condition. The equations and adjoint optimisation formulation depends only on the symmetry structure $\Gamma(\pi,\pi_{1},\pi_{2})$ which is determined from group theoretic integrals representing universal symmetry properties, and on the fluid parameters $(\rho, \mu)$ which appear as numerical parameters, and hence changing the fluid type only changes their numerical values, and not the form or applicability of the method. 
This method can in principle be used in the context of optimal control or machine learning models trained on extensive simulation datasets to quickly predict optimal adjustments without explicit re-solving.
The use of the Gelfand transformation is important because it explicitly encodes the symmetry information from the targeted molecular states directly into fluid velocity fields, enabling selective resonance. The structured coupling constants provide precise nonlinear interaction terms ($\Gamma$) in a structured form, which becomes critical for controlling the cascade evolution across scales. Also, the Gelfand transformed structured modes simplify the numerical analysis and iterative adjustments, clearly showing how modes at each cascade step evolve and couple. The Gelfand transform, as used here, enables therefore selective and iterative dynamical mode adjustment. 
In the standard chaotic thermalisation scenario, energy cascades down from a large scale $L$ to small scales (Kolmogorov length scale $\eta$). At scales near $\eta$, fluid kinetic energy is rapidly converted to random molecular motions, heating the fluid. This process is dominated by viscosity, resulting in random heat production 
\begin{equation}
E_{kin}\xrightarrow{viscous\;\;dissipation}Q_{heat}
\end{equation}
This process is generally irreversible, chaotic, and structureless. 
However, the scenario I am describing here differs fundamentally from this because the fluid modes are engineered to have specific symmetry and frequency matched precisely to discrete molecular vibrational states. Consider a toy model approximation of a molecular quantum oscillator 
\begin{equation}
H_{mol}=\frac{p^{2}}{2m}+\frac{1}{2}kQ^{2}
\end{equation}
where $Q$ is the vibrational amplitude (coordinate), $m$ is the effective mass, $k$ the force constant, and the vibrational frequency $\omega_{0}=\sqrt{k/m}$. Quantum mechanically, this mode has discrete energy levels
\begin{equation}
E_{n}=\hbar \omega_{0}(n+\frac{1}{2}),\;\; n=0,1,2,...
\end{equation}
For structured energy transfer to be efficient, the fluid mode frequency $\omega_{fluid}$ must match closely the molecular vibration frequency $\omega_{fluid}=\omega_{0}$. Under such a resonant condition, the coupling between the fluid and the molecular modes is enhanced dramatically. 
The fluid mode equation, with the structured mode $u$ is 
\begin{equation}
\rho\frac{du}{dt}+\rho\Gamma u^{2}+\mu |\omega|^{2}u=\rho f(t)-P(t)
\end{equation}
The quantum vibrational mode equation is 
\begin{equation}
m\frac{d^{2}Q}{dt^{2}}+m\gamma\frac{dQ}{dt}+m\omega_{0}^{2}Q=\alpha u(t)
\end{equation}
where $\alpha$ is a coupling strength, depending on symmetry match, $\Gamma$ is the structure constant, controlling structured fluid interaction, and $\gamma$ is the damping, which is a small quantum dissipation, or a vibrational relaxation rate. In resonance conditions $(\omega_{fluid}\sim \omega_{0})$, the energy coherently transfers from the structured fluid mode $u$ into the vibrational mode $Q$. Fast dissipation occurs if the energy moves rapidly into many random degrees of freedom. The dissipation timescale would be short. However, in the resonant coupling scenario, coherent resonance isolates a single molecular quantum mode, lengthening the timescale significantly. Mathematically, this is because the resonance condition leads to persistent coherent oscillations. Consider the resonance condition. The Fourier transform of the quantum mode equation is 
\begin{equation}
Q(\omega)=\frac{\alpha u(\omega)}{m(\omega_{0}^{2}-\omega^{2}+i\gamma\omega)}
\end{equation}
At resonance $\omega\sim\omega_{0}$ the denominator is minimal $|\omega_{0}^{2}-\omega^{2}|\ll 1$ making $Q(\omega)$ very large (high amplitude). This implies strong, persistent energy buildup in the molecular mode, rather than immediate randomisation. Eventually, even the resonantly excited mode will slowly transfer its energy to other molecular modes or phonons, resulting in thermalisation, but much more slowly and controllably, allowing coherent manipulation at molecular levels. Therefore, structured fluid modes persist and don't immediately vanish into random thermal motions because of resonant, symmetry-based coupling to specific discrete molecular vibrational modes. 
The structured fluid mode transfers energy more efficiently and rapidly to molecular vibrational modes than random thermalisation because the energy transfer rate under structured resonance conditions can be represented by a resonant coupling term:

\begin{equation}
\frac{d E_{vib}}{dt}\sim |\alpha u(\omega)|^{2}\delta(\omega-\omega_{vib})
\end{equation}
where $\alpha$ is the coupling strength, $u(\omega)$ is the structured fluid mode amplitude at resonance frequency $\omega_{vib}$ and the delta function emphasises the highly selective, frequency-specific nature of the coupling. In contrast, thermalisation involves random and broad-spectrum interactions, significantly slowing the rate of energy accumulation in any single mode.

\section{Numerical Simulation and Results}

To illustrate this approach, we conducted detailed numerical simulations focusing on the asymmetric vibrational mode of $CO_{2}$, a molecule of significant scientific and technological relevance. We considered a three-step nonlinear cascade, where the fluid frequency at each step is multiplied by a factor of approximately five, progressively bridging from the MHz to the THz range. The structured fluid perturbations were numerically solved using the generalised structured Navier–Stokes equations to identify resonance conditions clearly.

Our numerical results demonstrate a distinct and coherent buildup of vibrational energy in the $CO_{2}$ asymmetric stretch mode, evident from sharply defined resonance peaks in the frequency spectra obtained via Fourier analysis. We systematically explored the dependence of vibrational energy excitation efficiency on the nonlinear coupling constant , identifying optimal values that maximise energy transfer. Furthermore, spatial mode analysis revealed that structured fluid perturbations maintained their coherence through multiple cascade steps, confirming the viability of controlled, symmetry-driven energy cascades.
\begin{figure}[h!]
 \centering
  \includegraphics[width=0.5\textwidth]{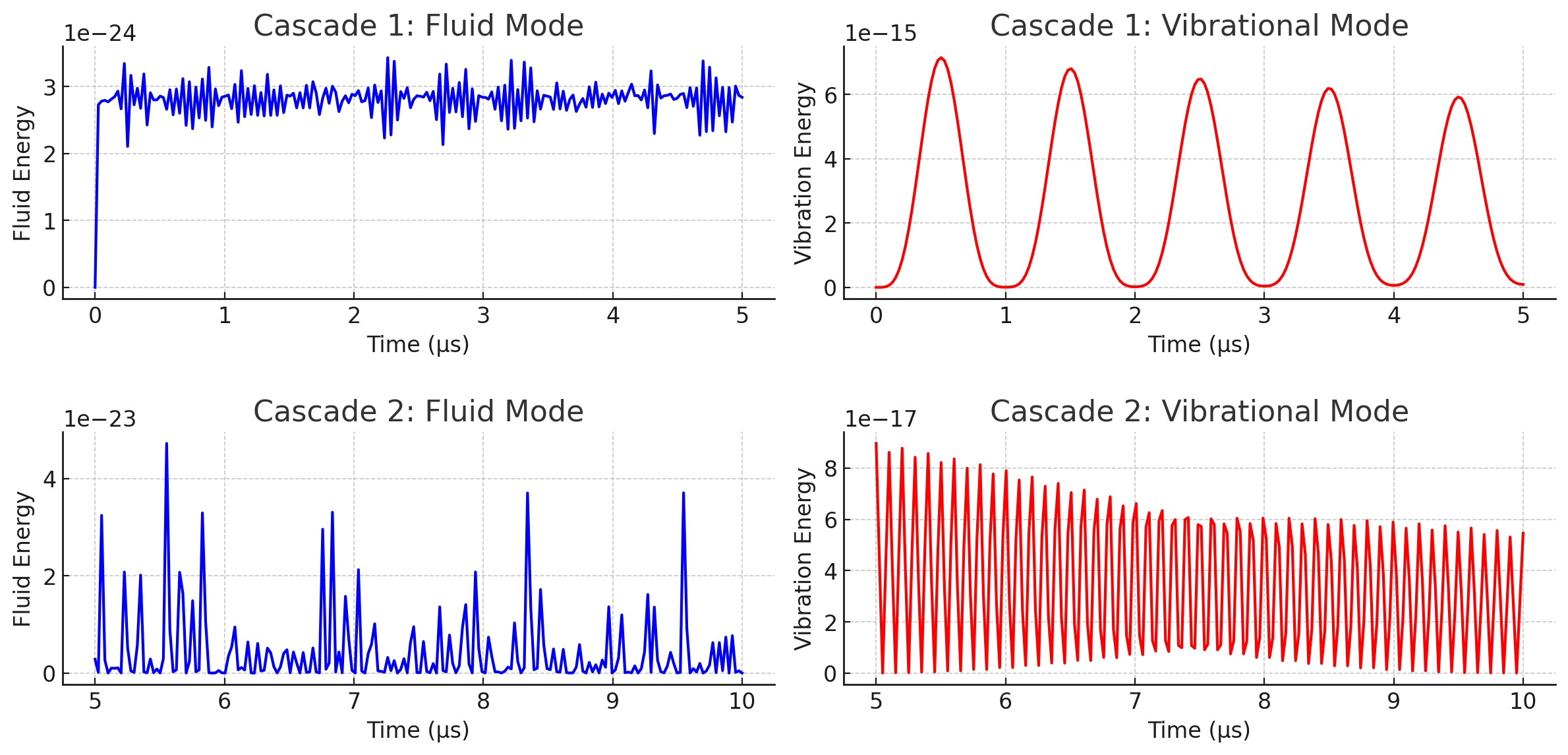}
 \caption{Second cascade step, highlighting the structured fluid perturbation and the targeted molecular vibrational mode}
 \end{figure}
Figure 1  presents a snapshot from the second cascade step, clearly demonstrating the selective enhancement of vibrational energy within the targeted molecular mode. The vibrational mode energy increases distinctly, indicating effective energy transfer from the structured fluid perturbation, as identified by sharply defined peaks in the frequency spectrum.
\begin{figure}[h!]
 \centering
  \includegraphics[width=0.5\textwidth]{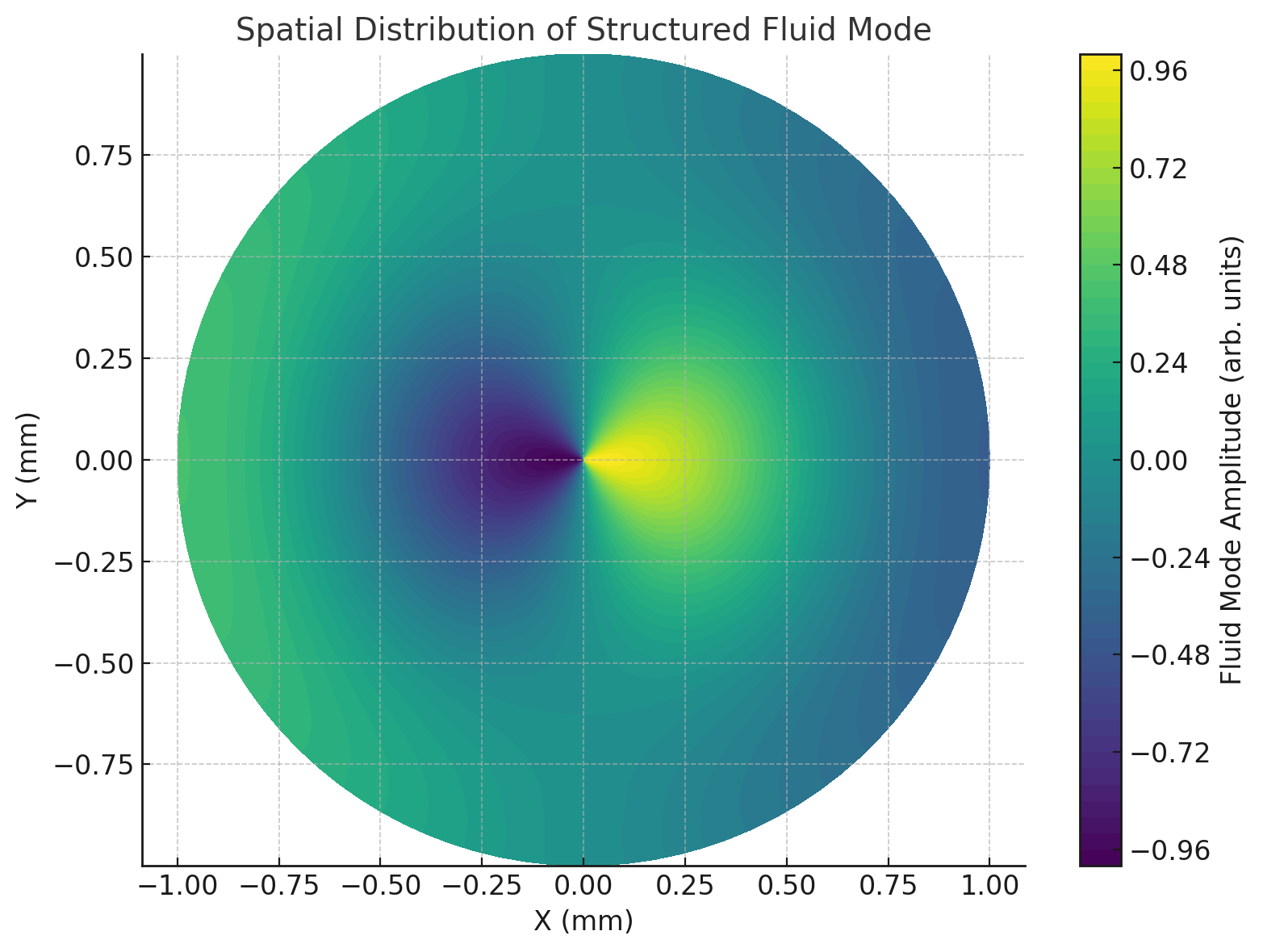}
 \caption{Spatial distribution of the structured fluid mode during an intermediate cascade}
 \end{figure}
Figure 2 illustrates the spatial distribution of the structured fluid modes during an intermediate cascade. The fluid perturbation exhibits coherent symmetry-adapted spatial patterns, designed to match the symmetry of the targeted molecular vibrational mode. The structured spatial coherence significantly accelerates energy transfer compared to random thermalisation processes.

The structured fluid mode transfers energy more efficiently and rapidly to the molecular vibrational modes than random thermalisation because structured interactions selectively target molecular resonance frequencies. This coherent interaction effectively bypasses intermediate random energy dispersal, maintaining a high level of energy coherence and significantly reducing the timescale required to achieve molecular excitation compared to conventional thermalisation processes.

\section{Conclusions and Perspectives}

Our results provide compelling evidence for the feasibility of coherent molecular vibrational excitation via structured fluid dynamics. The generalization of Navier–Stokes equations using symmetry-based Gelfand transforms opens unprecedented opportunities for precise control of molecular quantum states through fluidic environments. This approach significantly extends beyond conventional turbulence theory, offering practical pathways for experimental realization. Future work will focus on experimental verification of these theoretical predictions, optimization of fluidic devices for quantum state control, and exploration of broader applications in chemical synthesis, quantum information processing, and advanced energy technologies.

\end{document}